\documentclass[conference]{IEEEtran}
\IEEEoverridecommandlockouts
\usepackage{cite}
\usepackage{amsmath,amssymb,amsfonts}
\usepackage{algorithmic}
\usepackage{graphicx}
\usepackage{textcomp}
\usepackage{epstopdf}
\usepackage{xcolor}
\usepackage{bm}

\def\BibTeX{{\rm B\kern-.05em{\sc i\kern-.025em b}\kern-.08em
    T\kern-.1667em\lower.7ex\hbox{E}\kern-.125emX}}
\begin{document}

\title{Energy Saving in 5G and Beyond Networks: A Quantum Reinforcement Learning Approach\\
\thanks{This work is partially funded by Xenia Progetti srl under the Doctoral Program PNRR DM n. 117/2023, MUR.}
}

\author{
\IEEEauthorblockN{
Muhammad Usman\IEEEauthorrefmark{1},
Nguyen Van Huynh\IEEEauthorrefmark{2},
Marianna Lezzi\IEEEauthorrefmark{1},
Mariangela Lazoi\IEEEauthorrefmark{1}
}
\IEEEauthorblockA{\IEEEauthorrefmark{1}\textit{Department of Engineering for Innovation, University of Salento, Italy} \\
\IEEEauthorrefmark{2}\textit{School of Computer Science and Informatics, University of Liverpool, United Kingdom}\\
Emails: \{muhammad.usman, marianna.lezzi, mariangela.lazoi\}@unisalento.it \\
huynh.nguyen@liverpool.ac.uk}
}
\maketitle
\begin{abstract}
Energy saving has become a critical challenge in 5G and beyond networks. The rapid growth of connected devices has increased the overall network energy demand, driving operational expenditure to unsustainable heights. The Base Station (BS) accounts for the largest share of energy usage, typically consuming around 60-70\% of the Radio Access Network (RAN)'s total energy. Therefore, to address this issue, this article optimizes the BS's energy usage while accounting for the dynamic behavior of User Equipment (UE). Deep Reinforcement Learning (DRL) is a natural candidate for determining effective energy saving policies, such as automatically switching BSs on or off when user density is low or adjusting transmission power to balance energy efficiency and Quality of Service (QoS). However, its heavy training burden and the exponential growth of state and action spaces in dense 5G environments make exploration increasingly difficult. To overcome these limitations, we introduce a novel Quantum Reinforcement Learning (QRL) algorithm that leverages quantum principles, including superposition and entanglement, through parameterized quantum circuits, enabling significantly faster convergence than DRL, which relies on conventional deep neural networks. Extensive simulations demonstrate that the proposed QRL can substantially reduce energy consumption while maintaining QoS, even when UEs are highly dynamic and frequently switch their association with BS antennas. Additionally, QRL consistently outperforms DRL and Q-Learning in both convergence speed and learning complexity.
\end{abstract}


\section{Introduction} \label{intro}
The last decade has seen a rapid growth in mobile network traffic due to numerous improvements in 5G and beyond networks, including lower latency, higher data rates, and greater reliability \cite{1}. In this context, the spread of 5G technology has made the network dense. However, this staggering growth poses challenges for Mobile Network Operators (MNOs) in terms of Operational Expenditure (OpEx) and energy demand, which are expected to increase in the future \cite{3}.

Radio Access Network (RAN) alone accounts for about 60–80\% MNOs’ total energy \cite{5}, making RAN a major research focus. For energy-efficient 5G RAN, operators use a range of sleep-mode and cell-management strategies. These include mobility-based deactivation of underutilized small cells \cite{9} and Machine Learning (ML) techniques that optimize sleep modes by considering traffic, interference, and buffer conditions \cite{11}. Additional techniques include optimizing advanced sleep mode parameters, Base Stations (BSs) clustering schemes, and centralized dynamic sleep methods for energy saving. Furthermore, some studies update the sleep status of Radio Units (RUs) in real time according to network conditions, ensuring that only essential units remain active while preserving Quality of Service (QoS) \cite{18}.

Recent BS sleep modes reduce energy consumption, but suffer from reactivation delays that degrade QoS, shifting focus to ML-based optimization techniques. The problem is NP hard, and the dynamic behavior of User Equipments (UEs) adds more complexity \cite{19}, limiting the use of conventional ML due to its slow convergence. Deep Reinforcement Learning (DRL) performs better in dynamic environments \cite{20}, but its reliance on Deep Neural Networks (DNNs) slows learning in large, rapidly changing state spaces, hindering sub‑millisecond decisions. Quantum computing addresses these limits using superposition and entanglement to evaluate many states in parallel, enabling faster exploration and convergence \cite{8a}. This promises improved sample efficiency and lower training latency for real‑time wireless tasks. Through extensive simulations, we demonstrate that the proposed QRL approach achieves a significantly faster convergence rate compared to existing DRL techniques. It is important to note that the QRL framework can be implemented on classical computers using libraries such as Cirq. Our main contributions can be summarized as follows:
\begin{itemize}
    \item \textbf{Beam-selection with dynamic UE mobility:}  
    A novel framework is introduced using three-dimensional antennas spaced 120° apart, where BS transmit power is dynamically optimized as UEs update their positions and mobility patterns.
    \item \textbf{Quantum Reinforcement Learning for complexity reduction:}  
    To address the challenges of random UE mobility and per-antenna power allocation, parameterized quantum circuits are employed within a QRL framework, enabling efficient exploration of high-dimensional state spaces and faster convergence during policy training.
    \item \textbf{Joint optimization of throughput and energy efficiency:}  
    Unlike existing approaches \cite{8a, 32}, the proposed solution simultaneously manages network throughput and energy efficiency in dynamic 5G and beyond settings, ensuring QoS while improving sustainability.
\end{itemize} 
\section{System Model and Problem Formation} \label{SM}
In this section, we first present an overview of our considered system, followed by a description of the problem formation used in this framework.  
\subsection{Network Scenario}
\begin{figure*}[t]
    \centering
    \includegraphics[width=1.5\columnwidth]{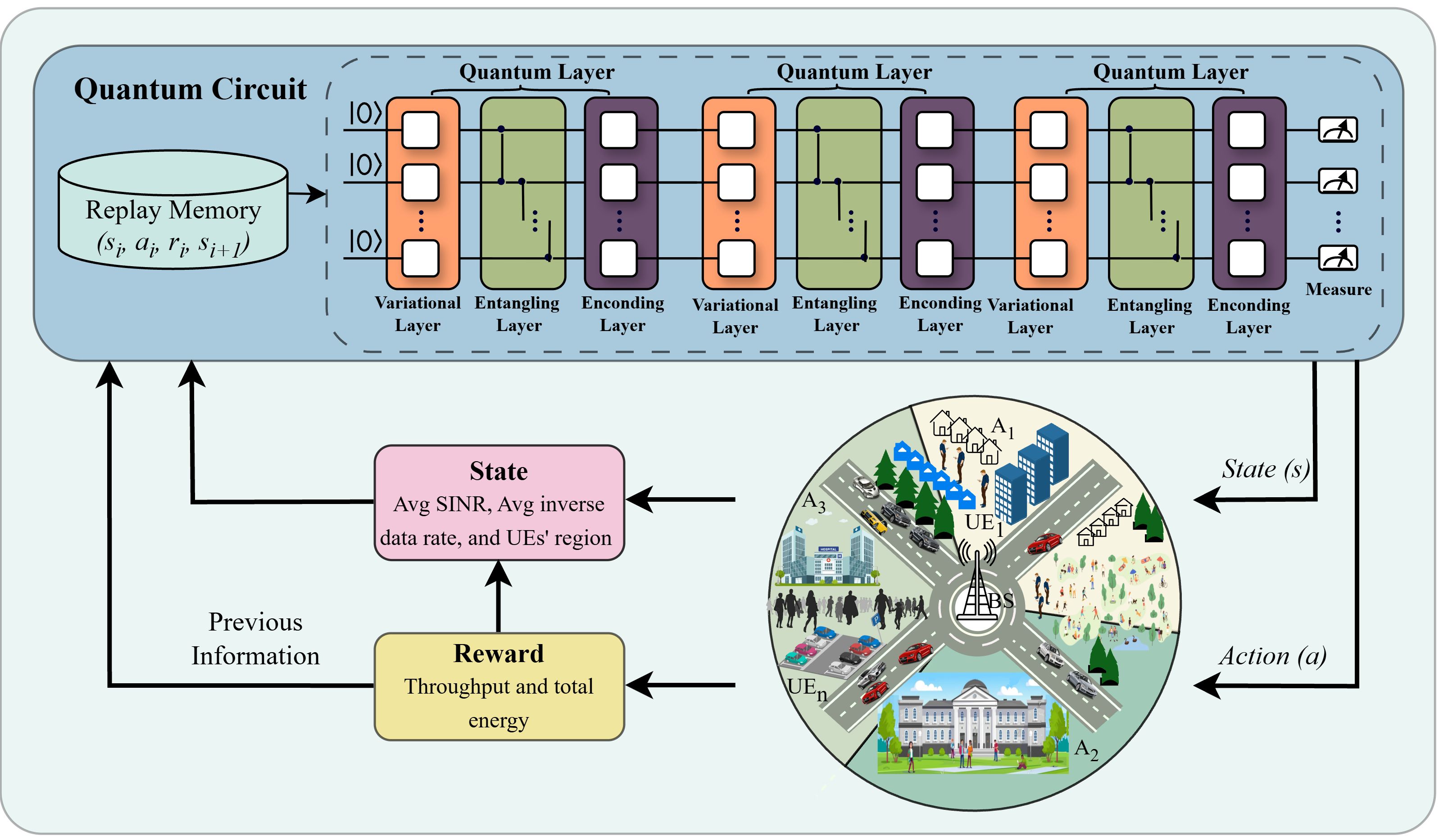}
    \vspace{- 1ex}
    \caption{Illustration of the system model. The BS consists of three antennas $(A_1, A_2, A_3)$ serving multiple UEs. Some UEs, for example, $UE_1$, are close to the antenna (requiring low energy), while $UE_n$ is farther away (requiring high BS transmission energy). UEs can move and connect to specific antennas. The environment provides state information (average SINR, inverse data rate, UE connectivity) to the QRL agent. The action space includes selecting an antenna and adjusting transmit power spectral density. The reward combines throughput and energy efficiency, with a penalty if the minimum data rate is violated, ensuring QoS is prioritized.}
    \vspace{-3ex}   
    \label{fig:SM}
\end{figure*}

The system model consists of a 5G BS with three antennas, spaced 120° apart, each covering a distinct sector. Users can move anywhere inside the BS’s coverage area and may receive signals from more than one antenna, as shown in Fig. \ref{fig:SM}. As the position of the user changes, the antenna association may also change. In sector overlap regions, a user connects to the antenna providing the strongest signal, but only if that antenna has enough spectrum resources. Once an antenna reaches full capacity, any new user is redirected to a different antenna, even if that antenna offers a lower data rate, until capacity becomes available again.
BS antennas can cooperate to reduce overall energy consumption. When only a few active users are in an antenna’s coverage area, that antenna may be deactivated to save energy. The users are then associated with a neighbouring antenna, which raises its transmission power, subject to allowable limits, to maintain service quality. A further power increase may be necessary to handle the additional traffic load. The goal is to sustain acceptable throughput and low packet loss while improving energy efficiency. In Fig. \ref{fig:SM}, for example, the BS uses three antennas covering regions $A_1–A_3$, and their transmission ranges adjust dynamically based on user distribution.

Let the set of UEs be denoted as $\boldsymbol{UE} = \{UE_1, UE_2, \dots, UE_n \}$, where each UE requires a specific data rate within bandwidth $B$. A BS requires less transmission power to serve nearby UEs (e.g., $UE_1$) and more power for distant UEs (e.g., $UE_2$), because Path Loss (PL) increases with distance. Therefore, BS antennas must dynamically adjust their transmission power based on each UE’s requirements to optimize energy consumption. Following \cite{35,36}, we assume the BS always allocates enough resources to serve all UEs; thus, packet loss and latency are outside this work’s scope. An RU is the smallest indivisible resource block, and multiple RUs may be assigned to a UE based on its required data rate. Each antenna has a finite number of RUs ($R_i$) to distribute, ensuring traffic demands are met while enabling efficient power management across antennas.

Practically, UEs are dynamic as they move over time and may connect to different antenna regions based on their proximity to the BS. Let $\boldsymbol{T} = \{t_1, t_2, \dots, t_n \}$ be the set of time steps, and let $P_{\mathrm{UE}}(t)$ denote the power consumed by a UE at time $t$. The BS position is fixed within each time slot. Let $\mathrm{A}$ be the set of antennas, and let $s_{i,t} \in {0,1,2}$ specify the operating state of antenna $A_i \in \mathrm{A}$: $0$ = off, $1$ = active, $2$ = wake‑up mode.

Since each UE connects to a specific antenna, the number of RUs that antenna must allocate to $UE_i$ (denoted $n_{\mathrm{UE}_{i,t}}$) to meet its required data rate is:

\begin{equation}
n_{\mathrm{UE}_{i,t}} = \left\lfloor \frac{\lambda_{\mathrm{UE}_{i,t}}}{r_{\mathrm{UE}_{i,t}}} \right\rfloor,
\end{equation}
where $\lambda_{\mathrm{UE}_{i,t}}$ is the traffic demand of $UE_i$ at $t$, $r_{\mathrm{UE}_{i,t}}$ is the data rate of $UE_i$ from the BS per RU, and is calculated as:

\begin{equation}
r_{\mathrm{UE}_{i,t}} = B \log_2 \left( 1 + \mathrm{SINR}_{\mathrm{UE}_{i,t}} \right),
\end{equation}
where the Signal-to-Interference-plus-Noise Ratio (SINR) for $UE_i$ at time $t$ is given by:

\begin{equation}
\mathrm{SINR}_{\mathrm{UE}_{i,t}} =
\frac{
P^{\mathrm{TX}}_{A_i,t} \, G_{A_i}^{\mathrm{tx}} \, G_{\mathrm{UE}_i}^{\mathrm{rx}} \, G^{\mathrm{MIMO}}
}{
PL^{\mathrm{LoS}}_{\mathrm{UE}_{i,t}} \, (I + N_0 B)
}.
\end{equation}

In this expression, $P_{A_i,t} ^{TX}$ is the transmission power of $A_i$ allocated to $UE_i, G_{A_i}^{tx}$ and $G_{UE_i}^{rx}$ are the antenna gains at the transmitter and receiver, and $G^{MIMO}$ is the MIMO gain. $I$ denotes the interference from other transmissions, and $N_0$ is the noise power spectral density, making $N_0B$ the total noise power over the bandwidth. The PL is either incorporated into the gains or explicitly modeled in the link budget.
The Line of Sight (LoS) PL for $UE_i$ at time $ t$ $PL_{UE_{i,t}} ^ {LoS}$ can be expressed as follows \cite{37}: 
\begin{equation}
PL^{\mathrm{LoS}}_{\mathrm{UE}_{i,t}}
= 16.9 \log_{10}\left(d_{{A_i},\mathrm{UE}_{i,t}}\right)
+ 32.8
+ 20 \log_{10}(f),
\end{equation}
where $d_{{A_i},{UE}_{i,t}}$ is the distance between $UE_i$ and its serving antenna $A_i$ at time $t$, and $f$ is the center frequency. For a BS, the total transmission power in time slot $t$ is then:
\begin{equation}
P_{\mathrm{BS}}^{\mathrm{TX}}
= P^{\mathrm{TX}}_{A,t}
\sum_{i \in \mathrm{UE}_{n,t}} n_{i,t}.
\end{equation}
Changing an antenna’s operational state adds control overhead, such as handling network and user data during activation and deactivation. The switching energy is then formulated as:
\begin{equation}
\label{SWE}
P^{\mathrm{SW}}_{{A_i},t} =
\begin{cases}
\eta \, |\mathrm{UE}_{i,t}|, & s_{i,t} = 2 \\[6pt]
\eta \, |\mathrm{UE}_{i,t-1}|, & s_{i,t-1} = 1,\; s_{i,t} = 0 \\[6pt]
0, & \text{otherwise},
\end{cases}
\end{equation}
where $\eta$ is the handover cost when a user moves between antennas of the same BS. We assume an antenna’s state does not change within a time slot, and an inactive antenna consumes no energy. The total energy consumption of antenna $A_i$ in time slot $t$, denoted $P_{{A_i},t}$, is then:
\begin{equation} \label{total_P}
P_{{A_i},t}
= s_{i,t}\left(P_{A_i}^{\mathrm{OP}} + P^{\mathrm{TX}}_{{A_i},t}\right)
+ |s_{i,t} - s_{i,t-1}| \, P^{\mathrm{SW}}_{{A_i},t},
\end{equation}
where $P_{A_i}^{OP}$ denotes the constant operational power of the antenna ${A_i}$, $P_{{A_i},t}^{TX}$ is the UE-dependent and adjustable transmission power, and $P_{{A_i},t}^{SW}$ is the switching power required to turn the antennas on or off. The total energy consumption of a BS can be calculated as,
\begin{equation}
P_{\mathrm{BS}_T}
= \sum_{i=1}^{A} \left( P_{A_i,t} + P^{\mathrm{idle}}_{A_i,t} + P^{\mathrm{SW}}_{Ai,t} \right),
\end{equation}
where $P_{{A_i},t}^{idle}$ represents the energy consumption of antenna ${A_i}$ in the idle state.
\subsection{Problem Formulation}\label{AA}
The problem is formulated for multi-antenna switching in a dynamic network scenario. The goal is to minimize the total energy consumption of the BS antennas over time $T$, given their transmission power and operating states $s_{i,t}$, while maintaining user throughput. Since switching antennas on/off incurs a cost, we avoid frequent switching.

Given a set of antennas, UEs, and their spatial‑temporal distribution over time $T$, the system must make three decisions at each time slot $t$. These include determining which antennas to activate based on UE demand, assigning UEs to antennas according to their mobility patterns, and selecting the transmission power level for each active antenna. For antenna $A_i$ and $UE_j$ at time $t$, we define the $s_{i,t}$, $P_{{A_i},t}^{TX}$, and association variable $x_{i,j,t} \in {0,1}$ (1 if $UE_j$ is served by $A_i$). The objective is to determine these variables for all $A_i \in A$, $j \in UE$, $t \in T$ that minimize the BS’s total energy consumption over $T$. The optimization problem is formulated as follows.

The optimization in Equ. \ref{optp} is NP-hard, and conventional solutions may be ineffective, which motivates the use of QRL. The first constraint (\ref{c1}) limits the number of RUs an antenna may assign based on its RU capacity. The second constraint (\ref{c2}) prevents a UE from associating with more than one antenna within the same time slot. The final constraint (\ref{c3}) restricts the transmission power of each RU at antenna $A_i$ to be within $P_{{A_i},min}^{TX}$ and $P_{{A_i},max}^{TX}$ for all ${A_i}\in A$.
\begin{align}
\min \quad & \sum_{t=1}^{T} \sum_{i=1}^{A} P_{{A_i},t} \label{optp} \\
\text{s.t.} \quad & \sum_{i \in \mathrm{UE}_{n,t}} n_{\mathrm{UE}_i,t} \leq R_i s_{i,t}, \quad \forall t \in T, \label{c1} \\
& \sum_{j \in \mathrm{Ant}} x_{a_i,j,t} s_{i,t} = 1, \quad \forall j \in \mathrm{UE}_n, \forall t \in T, \label{c2}\\
& P_{{A_i},\min}^{\mathrm{TX}} \leq P_{{A_i},t}^{\mathrm{TX}} \leq P_{{A_i},\max}^{\mathrm{TX}}, \quad \forall {A_i} \in A, \forall t \in T. \label{c3}
\end{align}
\section{QRL Approach for 5G and beyond Energy Saving}
The optimization problem is modeled as a Markov Decision Process (MDP) to capture the system’s dynamics and uncertainty. An MDP is a tuple $\left< S, \mathcal{A}, r\right>$, where $S$ is the state space, $\mathcal{A}$ the action space, and $r$ the immediate reward. Within this framework, BS antennas choose optimal actions (e.g., transmission power, coordination among antennas) based on current states like SINR and data rate, aiming to maximize their long‑term average reward.
\subsection{State}
The state captures the environment’s configuration. Over $T$ time slots, the system state is $\boldsymbol{S} = \{ \boldsymbol{s}_1, \dots, \boldsymbol{s}_T \}$, where the BS acts as an agent with its own observation space $O_t^j$. Each observation includes SINR, inverse data rate, and antenna connectivity of UEs. We define the global state as:
\begin{equation}
s_t^{j} = \left( O_t^{1},\, O_t^{2},\, \ldots,\, O_t^{UE} \right),
\end{equation}
where $O_t^{UE}$ is the set of observations, expressed as:
\begin{equation}
O_t^{j} = \left( \mathrm{SINR}_t^{j},\, \frac{1}{r_{\mathrm{UE}_i,t}^{\,j}},\, A_{i}^{j\,\mathrm{UE}_t} \right).
\end{equation}

\subsection{Action}
The actions in the time steps $t$ are $\boldsymbol{\mathcal{A}} = \{ \boldsymbol{{a}_1}, \dots, \boldsymbol{{a}}_T \}$. At each step, each UE selects an antenna, along with its transmission power, SINR, and data rate. The action at time $t$ is denoted as:
\begin{equation}
a_t^{j} = \left({A}_{i,t}^{j},\, P_{A_i,t}^{j,\mathrm{TX}},\, r_{\mathrm{UE}_i,t}^{\,j}, \, N_t^{j} \right),
\end{equation}
where $A_{i,t}^j$ is the antenna selected by the UE, $P_{A_i,t}^{j,\mathrm{TX}} \in {0,1,\ldots,\max(P_{A_i,t}^{j,\mathrm{TX}})}$ is the transmission power available to the UE, $r_{\mathrm{UE}_i,t}^{\,j}$ is the data rate, and $N_t^j \in {0,1,\ldots,N_{\max}}$ is the noise.
\subsection{Reward}
QRL uses a reward mechanism to guide each agent toward an optimal policy. We formulate the problem as a fully observable MDP that maximizes energy efficiency while preserving throughput for all UEs. At each time step $t$, the total throughput is computed as:
\begin{equation}
T_t = \sum_{\mathrm{UE}=1}^{N} r_{\mathrm{UE},t},
\end{equation}
and the total energy consumption from Equ. (7). To enforce QoS, a soft penalty is applied when the achieved throughput drops below the per‑UE data‑rate target. The resulting reward for agent $j$ is defined as:
\begin{equation}
R_t^{j} = \frac{T_t}{P_{\mathrm{BS}_T}} - \max\left(0,\; R_{\mathrm{QoS},\mathrm{UE}_n} - T_t \right),
\end{equation}
where $R_{QoS} = 5$ Mbps is the minimum data rate per UE. This formulation encourages agents to improve energy efficiency while keeping aggregate throughput above the QoS threshold.
\subsection{Quantum Optimization Aspect of MDP Problem}
QRL uses superposition and quantum parallelism. Unlike classical bits, quantum bits (qubits) have basis states $\lvert0\rangle$ and $\lvert1\rangle$ (Dirac notation) and can also exist in a superposition of both. By the superposition principle, a qubit state $\lvert\psi\rangle$ is a linear combination of the two basis states:
\begin{equation}
\lvert \Psi \rangle = \alpha \lvert 0 \rangle + \beta \lvert 1 \rangle,
\end{equation}
where $\alpha$ and $\beta$ are complex probability amplitudes for $\lvert0\rangle$ and $\lvert1\rangle$, with magnitudes that give the observation probabilities and satisfy the normalization condition.
\begin{equation}
\lvert \alpha\lvert^2 + \lvert \beta \lvert^2 = 1.
\end{equation}
The states $\lvert0\rangle$ and $\lvert1\rangle$ form an orthonormal Hilbert space basis. Applying a unitary operator to a superposition transforms all components simultaneously, producing a new superposition, a property known as quantum parallelism. Gates like Hadamard and phase gates generate and manipulate these superpositions. A networked system requires multiple qubits: $2^u$ states need $u$ qubits and corresponding multi‑qubit gates. The $u$-qubit state $\lvert \psi_u \rangle$ (base‑10) is a linear combination of orthonormal states:
\begin{equation}
\lvert \Psi \rangle = \sum_{i=0}^{2^{u}-1} r_i \lvert i \rangle.
\end{equation}
\subsection{Quantum State and Action Representation}
QRL improves action selection and reinforcement in classical RL using quantum computation. The collapse postulate enables probabilistic action selection, while amplitude amplification strengthens preferred actions, acting as the quantum counterpart of reinforcing high‑value actions. For each node $N_k$ at time $t_j$, the orthogonal quantum state $\lvert s_l^{k,j} \rangle$ and its corresponding action state $\lvert a_w^{k,j} \rangle$ form the quantized version of a classical RL state–action pair. The full set of discrete quantum states for node $k$ is
\begin{equation}
\boldsymbol{S}^{k} = \{\, \lvert \boldsymbol{s}_1^{k} \rangle,\, \lvert \boldsymbol{s}_2^{k} \rangle,\, \ldots,\, \lvert \boldsymbol{s}_{l}^{k} \rangle,\, \ldots \,\},
\end{equation}
and the corresponding action set available in state $\lvert s_l^{k,j} \rangle$ is
\begin{equation}
\boldsymbol{A}^{k}(\boldsymbol{s}_{l}^{k}) = \{\, \lvert \boldsymbol{a}_1^{k} \rangle,\, \lvert \boldsymbol{a}_2^{k} \rangle,\, \ldots,\, \lvert \boldsymbol{a}_{w}^{k} \rangle,\, \ldots \,\},
\end{equation}
If a node has $N_{st}$ eigenstates and $N_{ac}$ eigenactions, encoding them requires $u$ qubits for the state register and $v$ qubits for the action register, satisfying:
\begin{equation}
N_{\mathrm{st}} \le 2^{u} \le 2N_{\mathrm{st}}, \qquad
N_{\mathrm{ac}} \le 2^{v} \le 2N_{\mathrm{ac}}.
\end{equation}
The qubit‑level representation of a state and its action at time $t_j$ can be expressed as follows:
\begin{equation}
\lvert s_{l}^{k,j} \rangle 
= \lvert \mathrm{SINR}^{k,j},\, \frac{1}{r_{\mathrm{UE}}^{k,j}},\, A_{i}^{k,j} \mathrm{UE} \rangle
= \{\, \mathbf{q}_1,\, \mathbf{q}_2,\, \ldots,\, \mathbf{q}_u \,\},
\end{equation}
\begin{equation}
\lvert a_{w}^{k,j} (\, \lvert s_{l}^{k,j} \rangle \,) \rangle
= \lvert A_{i}^{k,j},\, P_{i}^{(k,j,\mathrm{TX})},\, N^{j} \rangle
= \{\, \mathbf{q}_1,\, \mathbf{q}_2,\, \ldots,\, \mathbf{q}_v \,\}.
\end{equation}
\subsection{QRL-Based State Value Updating}
In QRL, a unitary transformation $U$ maps the classical state $\lvert S^{k} \rangle$ to orthogonal eigenstates $\lvert s_{l}^{k,j} \rangle$. The transformed state is:
\begin{equation}
\lvert S^{k} \rangle = \sum_{l=1}^{N_{\mathrm{st}}} L_{l}\, \lvert s_{l}^{k,j} \rangle.
\end{equation}
When the agent uses $u$ qubits to encode the state space, quantum parallelism enables $2^u$ computational basis states to be evaluated simultaneously. Each basis state $\lvert s_{l}^{k,j} \rangle$ carries its own value estimate $V(s_l^{k,j})$, and all values are updated in parallel using the TD(0) rule.
\begin{equation}
V(s_{l}^{k,j})\leftarrow V\!(s_{l}^{k,j})
+ \zeta ( 
U^{k,j} 
+ \gamma\, V(s_{l}^{k,j+1})
- V(s_{l}^{k,j})
).
\end{equation}
A $u$-qubit register represents $2^u$ states simultaneously, so QRL updates an exponential number of states in a single operation. This quantum parallelism offers a theoretical speedup over classical RL, where updates are sequential.
\section{Performance Evaluation}
This section evaluates the QRL method against multiple baselines under various conditions. We first describe the configuration of the model parameters. We then analyze the simulation results, focusing on convergence, average power, and average throughput.
\subsection{Parameter Setting}
We model UEs randomly distributed within the coverage area; their mobility updates the UE–BS distance at each time step. The BS transmits 75 dBm per antenna, and noise is dynamically recalculated as UEs move. Bandwidth is 20 MHz at a 2 GHz carrier frequency. The BS uses four power levels (0\%, 25\%, 50\%, 100\%), defined based on the presence of the UE with respect to the BS. 

For DRL, we use standard hyperparameters from the literature. The network has two hidden layers (256 and 128 neurons) with tanh activation, a learning rate of 0.001, and the Adam optimizer. The $\in‑greedy$ exploration starts at $\in=1$ and decays to 0.01 with a factor 0.9999. The replay buffer holds 10,000 transitions, the target network updates every 5,000 steps, and the batch size is 32.

The QRL model uses three quantum layers, each with its own Adam optimizer to update parameters $\theta$, $\gamma$, and $\omega$. The circuit architecture follows \cite{32}, which forms the basis of our QRL framework. We evaluate QRL against three baselines by comparing transmit power and throughput. Throughput is computed from each antenna’s effective transmit power and the bits delivered per second to UEs. 
\begin{figure}[!ht]
    \centering
    \vspace{-3ex}
    \includegraphics[width=0.9\columnwidth]{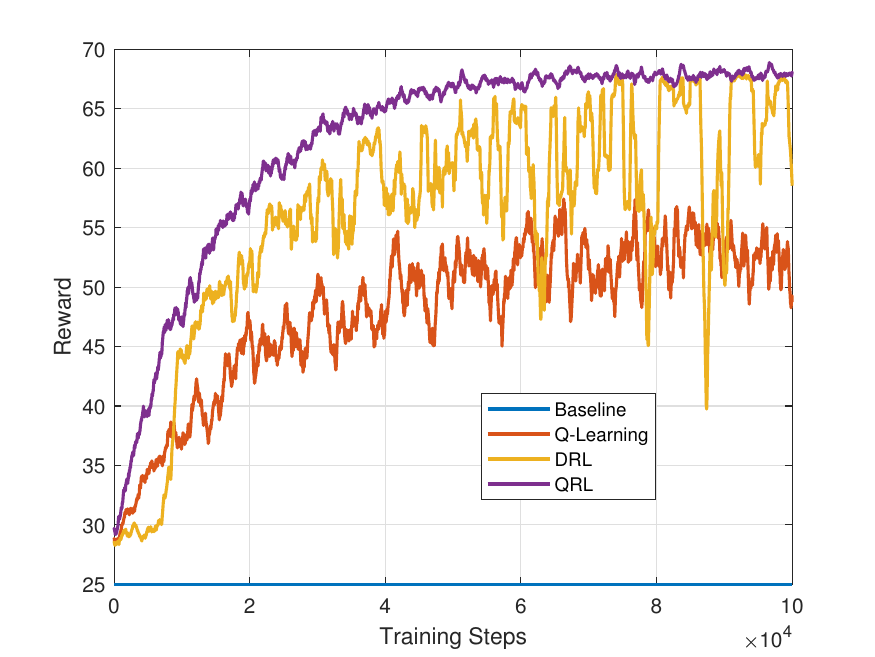}
    \vspace{-3ex}
    \caption{Reward Progression of the Models with Training Steps.}
    \vspace{-2ex}   
    \label{fig:train}
\end{figure}
\subsection{Simulation Results}
\begin{figure}[!ht]
    \centering
    \vspace{-3ex}
    \includegraphics[width=0.9\columnwidth]{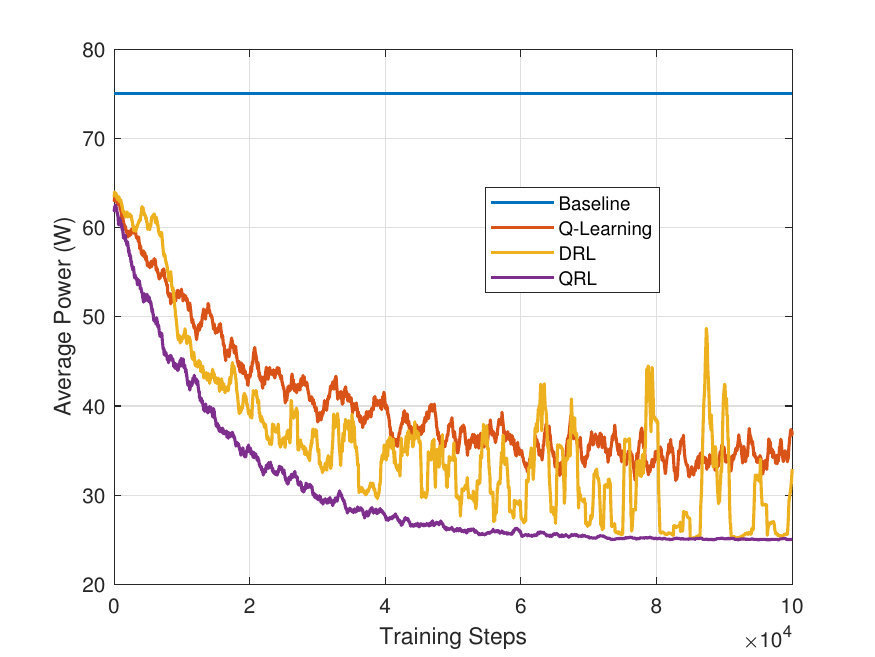}
    \vspace{-3ex}
    \caption{Power Consumption During Training.}
    \vspace{-3ex}   
    \label{fig:P_con}
\end{figure}

Fig. \ref{fig:train} compares the convergence speed of QRL against DRL and Q‑learning.
QRL converges faster and more stably, reaching an average reward of approximately 65 within the first 6,000 steps, whereas neither baseline model achieves comparable performance even after 100,000 steps across repeated runs. This demonstrates the strong learning capability of our quantum circuit architecture, enabled by quantum superposition for environment exploration.
Fig. \ref{fig:P_con} shows how power consumption evolves during training for Q‑learning, DRL, and QRL. 
The baseline maintains constant power, while QRL rapidly reduces power, outperforming both DRL and Q‑learning. Q‑learning initially matches DRL, but DRL becomes more energy‑efficient after about 10,000 steps, with the two alternating thereafter. QRL consistently achieves the lowest power and converges much faster than the others.

\begin{figure}[!ht]
    \centering
    \vspace{-3ex}
    \includegraphics[width=0.9\columnwidth]{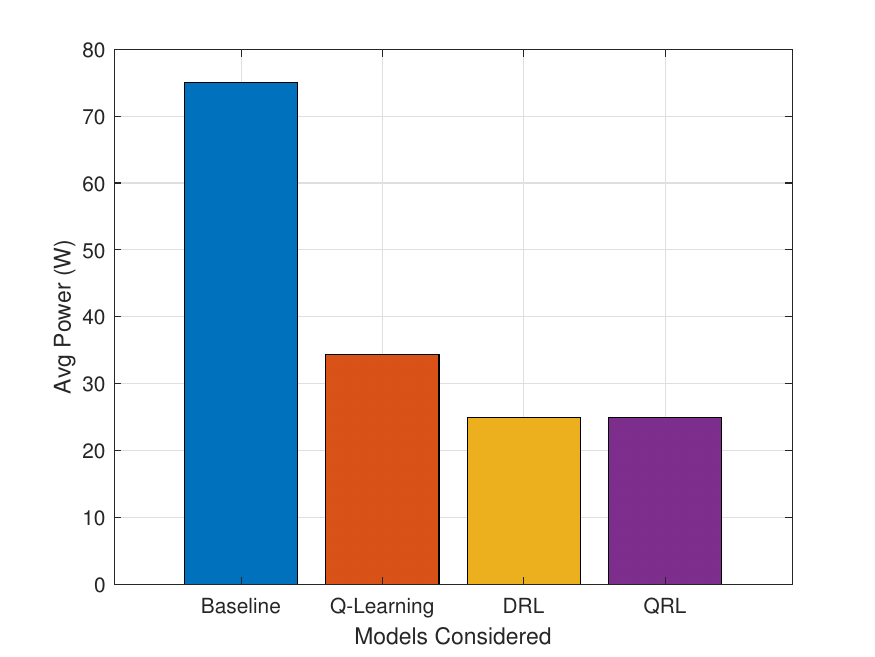}
    \vspace{-3ex}
    \caption{Average Power of BSs.}
    \vspace{-2ex}   
    \label{fig:avg_p}
\end{figure}
\begin{figure}[t]
    \centering
    \vspace{-3ex}
    \includegraphics[width=0.9\columnwidth]{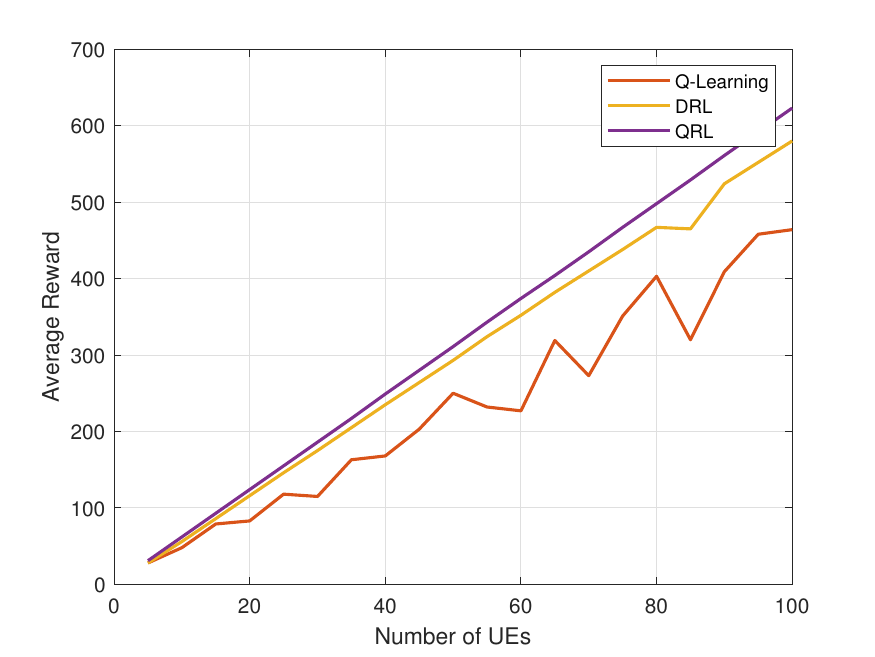}
    \vspace{-3ex}
    \caption{Average Reward with different numbers of UEs.}
    \vspace{-4ex}   
    \label{fig:reward_vs_UEs}
\end{figure}

Fig. \ref{fig:avg_p} shows the mean power consumption of the BS antennas. DRL and QRL operate at similar average power levels,
\begin{figure}[!ht]
    \centering
    \vspace{-3ex}
    \includegraphics[width=0.9\columnwidth]{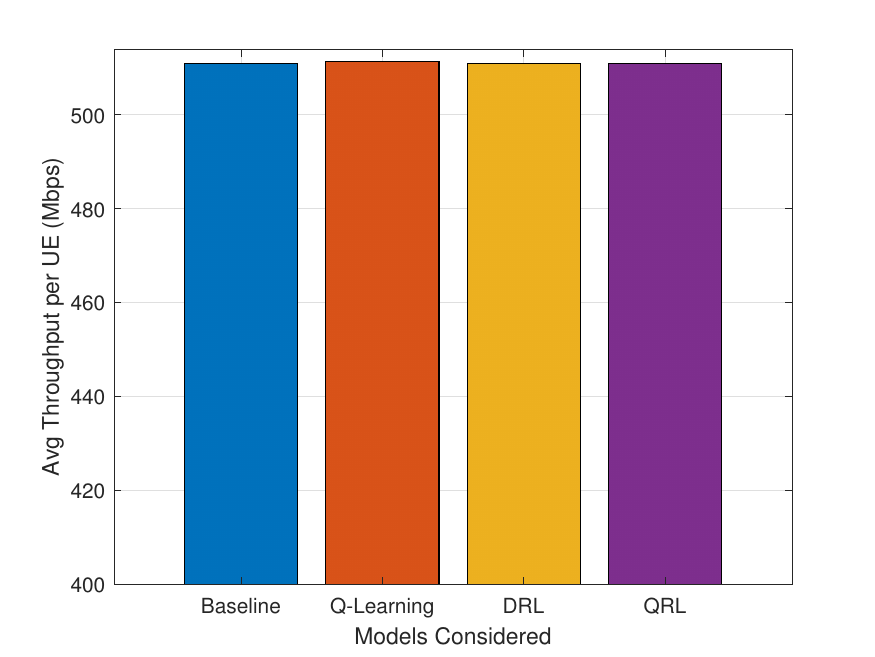}
    \vspace{-3ex}
    \caption{Average Throughput per Dynamic UEs.}
     \vspace{-2ex}   
    \label{fig:avg_thr}
\end{figure}
 but DRL converges much more poorly than QRL. This is largely due to the low UE density, where a minimum transmit power of 25\% is needed to maintain QoS.
The energy optimization strategies do not degrade performance, as average throughput remains stable (Fig. \ref{fig:avg_thr}). This confirms that reducing BS energy consumption does not harm UE throughput.
\begin{figure}[!ht]
    \centering
    \vspace{-3ex}
    \includegraphics[width=0.9\columnwidth]{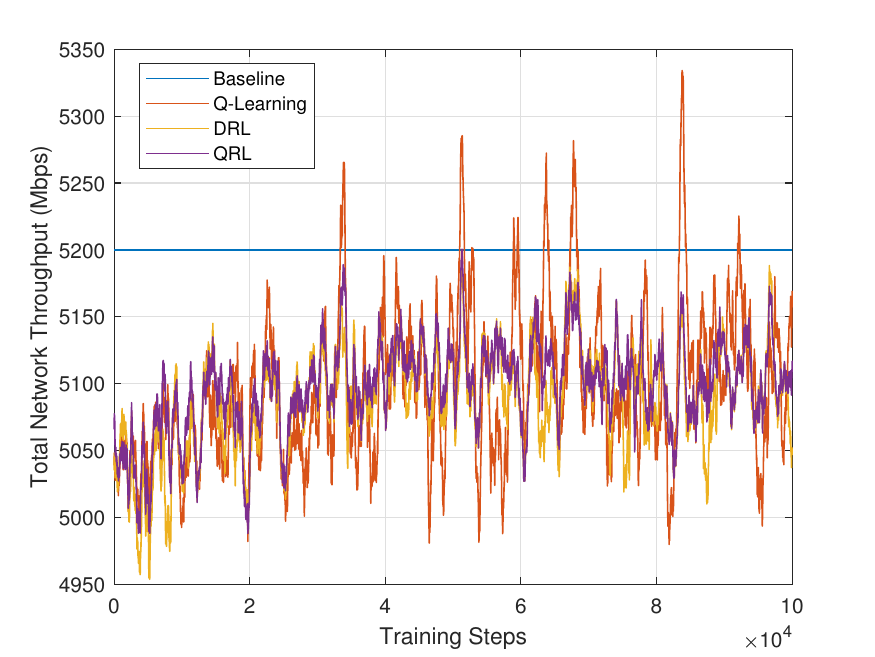}
    \vspace{-3ex}
    \caption{Throughput During Training.}
    \vspace{-2ex}   
    \label{fig:thr-tstp}
\end{figure}
Fig. \ref{fig:thr-tstp} shows that QRL has fewer throughput fluctuations than Q‑Learning and DRL, thanks to its better convergence and ability to maintain QoS during mobility. Q‑Learning performs worst, with rapid fluctuations due to poor stability. DRL is better, but QRL performs best, keeping throughput close to the baseline.
Fig. \ref{fig:reward_vs_UEs} compares models as the number of UEs increases from 0 to 100 under the same conditions. Q‑Learning shows large fluctuations and lower rewards, with instability growing as UEs increase. DRL performs more smoothly and achieves higher rewards, with smaller fluctuations indicating better stability. QRL achieves the highest and most consistent rewards, with an almost linear, stable trend, demonstrating strong convergence and scalability. Thus, increasing the number of UEs does not affect performance, making the model suitable for ultra‑dense networks. 
\section{Conclusion}
In this paper, we propose an energy saving strategy for 5G and beyond networks using Quantum Reinforcement Learning (QRL) to manage dynamic User Equipment (UEs). The high energy demand of 5G networks, driven by unpredictable user behaviour, increases Operational Expenditure (OpEx) for mobile operators. Optimising energy usage is therefore critical, especially in dense deployments. However, the optimal Base Station (BS) antenna transmission power configuration is difficult to identify due to the dynamic and uncertain nature of UEs. To address this, we developed a QRL agent that uses a parameterized quantum circuit (with up to 8 qubits and 3 layers). Simulations using a custom multi-user environment show that our QRL approach reduces BS energy consumption while maintaining QoS and requires significantly fewer training steps than classical DRL. Although the current simulator-based implementation uses a simple network model with a few UEs and a single BS, in the future we plan to extend the work to multiple BSs and a denser network architecture.

\end{document}